# Bias Dependent Variability of Low-Frequency Noise in Single-Layer Graphene FETs

Nikolaos Mavredakis*, Ramon Garcia Cortadella, Xavi Illa, Nathan Schaefer, Andrea Bonaccini Calia, Anton Guimerà-Brunet, Jose Antonio Garrido and David Jiménez

*Abstract*— **Low-frequency noise (LFN) variability in graphene transistors (GFETs) is for the first time researched in this work. LFN from an adequate statistical sample of long-channel solution-gated single-layer GFETs is measured in a wide range of operating conditions while a physics-based analytical model is derived that accounts for the bias dependence of LFN variance with remarkable performance. It is theoretically proved and experimentally validated that LFN deviations in GFETs stem from physical mechanisms that generate LFN. Thus, carrier number ($\Delta N$) due to trapping/detrapping process and mobility fluctuations ($\Delta \mu$) which are the main causes of LFN, define its variability likewise as its mean value. $\Delta N$ accounts for an M-shape of normalized LFN variance versus gate bias with a minimum at the charge neutrality point (CNP) as it was the case for normalized LFN mean value while $\Delta \mu$ contributes only near the CNP for both variance and mean value. Trap statistical nature is experimentally shown to differ from classical Poisson distribution at silicon-oxide devices, and this is probably caused by electrolyte interface in GFETs under study. Overall, GFET technology development is still in a premature stage which might cause pivotal inconsistencies affecting the scaling laws in GFETs of the same process.**

## I. INTRODUCTION

Nowadays, the limitations on advanced CMOS technologies and the predictions for deceleration of Moore's law, has led both the scientific community and semiconductor industry to turn their attention at the development of emerging technologies based on 2-Dimensional (2D) materials such as graphene.[1, 2] Graphene's extraordinary characteristics such as carrier mobilities up to *2.10⁵* cm²(Vs)⁻¹ and saturation velocities of *4.10⁷* cm(s)⁻¹ have placed graphene transistors (GFET) as an quintessential prospective for future applications.[3] Despite the fact that the lack of bandgap in graphene due to its semimetal nature makes single-layer (SL) GFETs unsuitable for digital electronics, there has been an enormous increase of analog and RF circuits designed with

GFET technology such as: frequency multipliers,[4, 5] voltage control[6, 7] and ring oscillators[8] as well as terahertz detectors.[9-11] Moreover, GFETs are also widely used and tremendously improve the performance of chemical-biological sensors and optoelectronic devices.[12-17]

It is not enough for recently developed GFET technologies just to exhibit optimal performance, but they should also be reliable and consistent with CMOS ones since the majority of industry aims to develop new GFET processes based on their pre-existing infrastructures designed for silicon devices.[18, 19] Variability issues are of outmost importance in advanced semiconductor technologies. Regarding GFETs, the thorough study of such effects is crucial for the transition from immature technologies and prototyped devices mainly fabricated in research labs to large-scale wafer production which will lead to a boost of graphene-based applications and products. There are two sources of variabilities in graphene: a) environmental effect variabilities such as interface traps and b) material imperfection variabilities such as edge disorders.[18] In this work we focus on low-frequency noise (LFN) variability for SL GFETs mainly derived from interface trap statistics which, according to our knowledge, is for the first time investigated. LFN's contribution to the aforementioned circuits is very crucial as it can be upconverted to deleterious phase noise in such high frequency applications.[4-11] In addition, it can affect the sensitivity of sensors[12-17] while LFN deviations can also be proved useful for such sensing applications.[18] Moreover, LFN examination can provide crucial information as far as the quality of the devices and their interfaces is concerned.[20, 21]

Three main effects are considered responsible for the generation of LFN in semiconductor devices and consequently GFETs; firstly the carrier number fluctuation mechanism ($\Delta N$), secondly the mobility fluctuation mechanism ($\Delta \mu$) and finally the series resistance ($R_c$) contribution ($\Delta R$). $\Delta N$ model stems from the trapping/detrapping process at semiconductor devices.[22, 23]

N. Mavredakis and D. Jimenez are with the Departament d'Enginyeria Electrònica, Escola d'Enginyeria, Universitat Autònoma de Barcelona, Bellaterra 08193, Spain. (e-mail: Nikolaos.mavredakis@uab.es).

R. G. Cortadella, Nathan Schaefer, A. Bonaccini C. and J. A. Garrido are with the Catalan Institute of Nanoscience and Nanotechnology (ICN2), CSIC, Barcelona Institute of Science and Technology, Campus UAB, Bellaterra, Barcelona, Spain.

X. Illa and A. Guimera-Brunet are with the Instituto de Microelectronica de Barcelona, IMB-CNM (CSIC), Esfera, UAB, Bellatera, Spain and the Centro de Investigacion Biomedica en Red en Bioingenieria, Biomateriales y Nanomedicina (CIBER-BBN), Madrid, Spain.

J. A. Garrido also is with ICREA, Pg. Lluis Companys 23, 08010 Barcelona, Spain.

In more detail, a free carrier can be captured by an active trap near the dielectric interface and within a few $kT$ from the Fermi level and then emitted back at the conduction path, and as a result a Random Telegraph Signal (RTS) is generated which corresponds to a Lorentzian Power Spectral Density (PSD). For transistors with channels larger than a few hundred nanometers, the number of active traps is high and consequently the superposition of the generated Lorentzian spectra can result in LFN PSDs inversely proportional to frequency under the condition that the traps are uniformly distributed. This is also known as 1/f noise and was first proposed by McWhorter.[24] Minimization of device dimensions in advanced CMOS technology nodes has led both the LFN mean value and variance to be dominated by RTS[25-27] but this is not yet the case in GFETs. $\Delta\mu$ model occurs due to fluctuations of the bulk mobility and is described by the empirical Hooge equation[28] while $\Delta R$ one is caused by $R_c$ contribution especially at high current regimes. Several physics-based models, simpler or more analytical ones, are available in literature describing both LFN mean value[29-33] and variance[34-40] in CMOS transistors. Most of the LFN variance models focus on its area dependence for short channel CMOS devices where RTS prevail[34-37] while bias dependence is also analyzed in some of them.[38-40] Characterization and modeling of the standard deviation of the natural logarithm of LFN is also very common in bibliography since LFN deviations follow a log-normal distribution.[25, 27, 35-37, 39-40]

A lot of research has also been conducted regarding LFN in GFETs[41-55] and the findings agree that the same mechanisms ($\Delta N$, $\Delta\mu$, $\Delta R$) are responsible for the generation of LFN. In fact, it has been shown that the nature of LFN in GFETs strongly relies upon the number of layers of the device.[47] In transistors with many layers, volume noise ($\Delta\mu$) prevails while the fewer the layers the more significant the surface LFN ($\Delta N$) becomes. In this work, SL GFETs are governed by trapping/detrapping mechanism which causes an M-shape gate-bias dependence of output LFN divided by squared drain current ($S_{ID}f/I_D{}^2$), referred to $1$ Hz, with a minimum at the charge neutrality point (CNP). Residual charge, which dominates at the CNP, can proportionally increase the LFN minimum. Similarly, non-homogeneous charge density at the channel, caused by a relative high drain voltage, can also increase the LFN minimum.[53] $\Delta\mu$ model can also contribute to LFN near the CNP always with a $\Lambda$-shape trend even at SL GFETs while $\Delta R$ has been observed at higher current regime where $R_c$ is important.[53-55] LFN in GFETs can be reduced after electron-beam irradiation[49] while the same can be achieved with the usage of substrates such as boron nitride.[50, 51] A number of models have been proposed to describe the behavior of LFN mean value on GFETs but most of them[43, 45, 46] are based on a simple approximation for $\Delta N$ model taken from CMOS devices[26, 29, 31] which introduces an $\sim(g_m/I_D)^2$ trend of $S_{ID}f/I_D{}^2$ LFN. The latter approach can only be functional under uniform channel conditions.[54, 55] Recently, a complete physics-based LFN model was proposed[53-55] which is valid in all operating regions since it accounts for all non-homogeneities of the device.

While there is a significant number of works regarding LFN mean value modeling in GFETs, no studies are available that deal with LFN variability in these devices even though it is equally significant to its mean value. The large deviations of LFN observed in GFETs, makes it urgent to develop statistical LFN models to investigate the physics behind this variability. In the present work, such an analysis that combines both the experimental characterization of LFN variance data and the derivation of a new physics-based statistical compact model, is proposed for the first time. The proposed model reveals the relation of LFN variability with the fundamental physical generators of LFN in GFETs ($\Delta N$, $\Delta\mu$). As it is known from CMOS technology, LFN variance is connected with operating conditions in larger devices[39, 40], which is the case for the GFETs under study in the present analysis. The proposed model is based on the recently established chemical-potential based one regarding LFN mean value[53] as well as CV-IV behavior.[56-58] (See Section S.1 and Figure S1 in the Supporting Information for more details on the CV-IV-LFN model.) The principal idea for the LFN mean value model calculation was to divide the device channel into microscopic uncorrelated local noise sources. The local noise PSD that originates from each LFN mechanism was then calculated and integrated from Source to Drain in order to analytically evaluate the contributions from $\Delta N$, $\Delta\mu$ and $\Delta R$ LFN; by adding these contributions the total LFN PSD is obtained.[33] Similarly, in the present work, the variance is calculated locally for each LFN mechanism by applying fundamental laws of statistics.[40] Then, the integration along the channel leads to an analytical compact solution with the help of the chemical-potential based model mentioned above. The derived model describes very accurately the experimental data and similarly to CMOS devices,[39-40] $\Delta N$ and $\Delta\mu$ variance present a bias dependence similar to their means. Thus, $\Delta N$ variance is responsible for an M-shape of total LFN variance centered at the CNP, while $\Delta\mu$ variance is more significant near the CNP. The proposed LFN variance compact model can be easily implemented in Verilog-A and annexed at the chemical-potential based model mentioned above[53-58] and then included in circuit simulators.

Number of traps are known to follow a Poisson distribution in silicon-oxide devices[27, 35-38]. However, solution-gated (SG)-SL GFETs, which are examined in the present work, exhibit some special characteristics due to their liquid interface.[43, 59-63] The trapping/detrapping mechanism that generates LFN can occur either near the surface of the polyimide substrate or at the graphene-electrolyte interface where processes of association/dissociation with charged moieties can take place.[60-63] In the present study, the proposed physics-based compact model of LFN variability is used to extract conclusions about the statistics of charge traps in these devices, demonstrating for the first time that they do not follow a Poisson distribution.

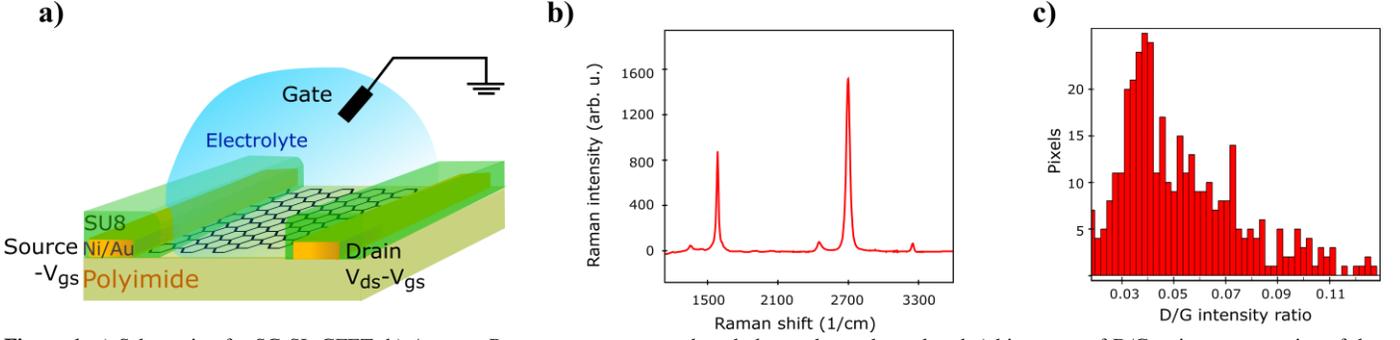

**Figure 1.** a) Schematic of a SG-SL GFET. b) Average Raman spectrum over the whole graphene channel and c) histogram of D/G ratio representative of the defects density distribution.

## II. Results-Discussions

On wafer LFN and IV measurements were conducted at top SG-SL GFETs[59] at three different device geometries: a) *W=100* μm/*L=100* μm, b) *W=50* μm/*L=50* μm and c) *W=20* μm/*L=20* μm where *W*, *L* are the width and length of the device respectively. Top gate voltage was swept from strong p-type to strong n-type region including the CNP with a step of *5* mV while drain voltage was constant, $V_{DS}=50$ mV. Arrays of SG-SL GFETs were fabricated (See Experimental Section for more details on fabrication and measurements procedures.) and thus, a significant number of samples were measured for each available geometry in order to have adequate data to characterize LFN statistics. In more detail, the analysis took place at around *48-50* samples from geometry (a), *25-28* samples from geometry (b) and *23-25* samples from geometry (c) after the exclusion of outliers. The reason why the number of samples is not constant for each geometry is that some measurements might behave as outliers in some specific operating conditions but not in the whole range. While $I_D$ and $S_{ID}$ were measured for *100* top gate voltage values, only *11* of them were chosen for the LFN variability analysis in order to speed up the process. These values were extended from high- to low-current regime both at p- and n-type regions in order to permit the thorough study of the LFN variance at all the operating conditions. The schematic of the device under test is shown in **Figure 1a** where the graphene channel, the metal contacts, the SU8 passivation, the electrolyte gate and the reference electrode are shown. Figure 1b illustrates the average Raman spectrum from *400* points in the *W=20* μm/*L=20* μm GFET area after the transfer of graphene (see Methods section). Figure 1c shows an histogram representing the D/G ratio for each of the measured spectra, indicating the rather low density of defects in the graphene lattice.

*IV Model Validation:* The first step towards the modeling of LFN mean and variance is the extraction of physical parameters related to their stationary response. In **Figure 2a**, IV model is validated[56-58] for the three GFET geometries under test. The transfer characteristics (drain-source current $I_D$ vs. effective gate voltage $V_{GEFF}$) are shown for all regimes of operation, near and away the CNP, with $V_{GEFF}$ calculated as the gate voltage ($V_G$) minus the voltage at the CNP ($V_{CNP}$). Figure 2a also presents the fitting from the chemical-potential based model,[56-58] showing a close match with the experimental data. The fundamental parameters of the IV model such as mobility $\mu$, top gate capacitance $C_{top}$, flat band voltage $V_{GSO}$, residual charge $\rho_0$ and contact resistance $R_c$ are extracted and presented in Table 1. Derived parameters from every GFET are very close, apart from $\mu$ and $V_{GSO}$ which are quite heightened for the larger device, which is indicative of the elevated $I_D$ data observed there. Also, $R_c$ for the medium sized device is a little decreased.

*LFN Mean Value model:* As described before, $\Delta N$, $\Delta \mu$, $\Delta R$ are the main generators of LFN in GFETs. Regarding LFN variance, the first two are going to be investigated in this work. LFN normalized by the area over squared drain current ($WLS_{ID}f/I_D^2$), at *1* Hz, is widely used in literature for the study of LFN variance. The reason for using this normalization is that the variance of this term presents a $\sim1/(WL)$ dependence.[35-36, 39-40] In Figure 2b, this form of depiction of LFN data is experimentally shown to follow a log-normal distribution for the first time in GFETs. Cumulative distribution function (CDF) of the natural logarithm of $WLS_{ID}f/I_D^2$ for the *100* μm/*100* μm GFET is presented at three different $V_{GEFF}$. Figure 2c indicates the $WLS_{ID}f/I_D^2$ variability for the same device, which is much higher than and uncorrelated with the $V_{CNP}$ variability. This observation proves that LFN variance is not related to variability of the IV characteristics but mostly connected with the physical mechanisms that generate LFN. (for the equivalent plots as Figure 2b and 2c for the rest of the GFET areas see Figure S2 at Section S.2 of the Supporting Information.) Therefore, for the derivation of the LFN variance model, it is crucial to first determine the parameters of the LFN mean value model that are sensitive to variations. These are the number of traps $N_{tr}$ from $\Delta N$ effect and Hooge parameter $\alpha_H$ from $\Delta \mu$ one. The $WLS_{ID}f/I_D^2$ PSD locally in the device's channel for a slice $\Delta x$, for both mechanisms, is given by:[53]

$$\left. \frac{S_{\delta I_D}}{I_D^2} \right|_{\Delta N} Wlf = \left( \frac{e}{Q_{gr}} \frac{C_q}{C_q + C} \right)^2 \frac{N_{tr}}{W \Delta x} \tag{1}$$

$$\left. \frac{S_{\delta I_D}}{I_D^2} \right|_{\Delta \mu} Wlf = \frac{e}{Q_{gr}} \frac{WL\alpha_H}{W \Delta x} \tag{2}$$

respectively, where $N_{tr}=WLN_t=KT\lambda N_T$[31, 32] is the number of active traps, $N_t$ is the trap density in (cm$^{-2}$) and $N_T$ is the volumetric trap density in (eV$^{-1}$cm$^{-3}$) which is used as a LFN mean value model parameter.[53] $K$ is the Boltzman constant, $T$ is the absolute temperature, $\lambda\approx0.1$ nm is the tunneling attenuation distance, $e$ is the electron charge, $Q_{gr}$ is the graphene charge stored in the quantum capacitance $C_q$,

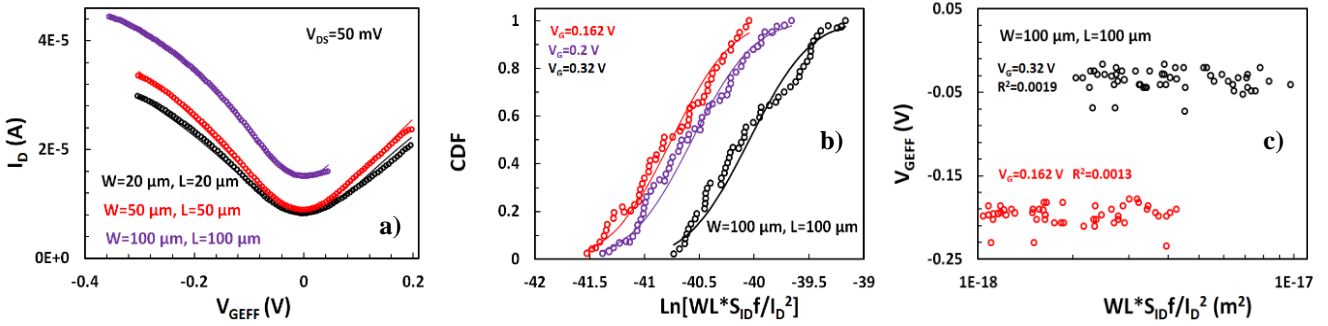

**Figure 2.** a) Drain current $I_D$ vs. top gate voltage overdrive $V_{GEFF}$, for GFETs with $W/L=20$ μm/20 μm, $50$ μm/50 μm and $100$ μm/100 μm. b) Cumulative distribution function (CDF) of natural logarithm of normalized LFN $Ln(WLS_{ID}f/I_D^2)$, referred to $1$ Hz, for GFET with $W/L=100$ μm/100 μm, shows a log-normal distribution. c) Variability of $WLS_{ID}f/I_D^2$, referred to $1$ Hz, is much higher and uncorrelated with variability of $V_{GEFF}$ for GFET with $W/L=100$ μm/100 μm.

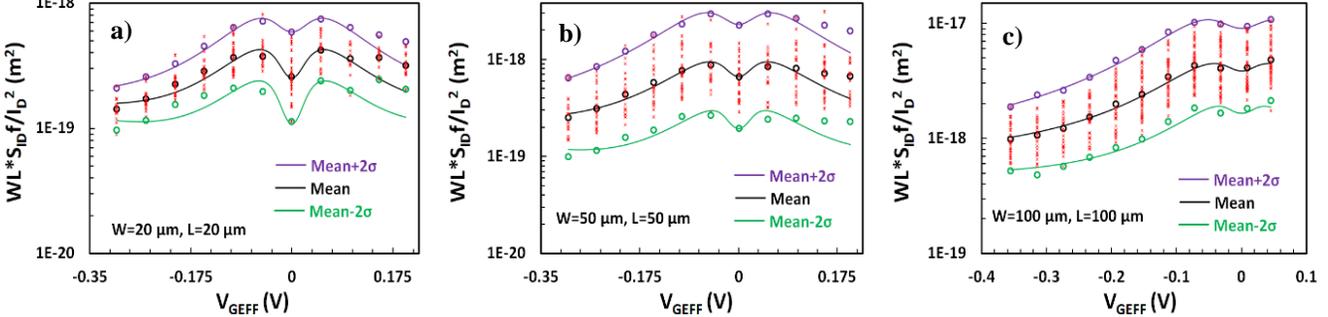

**Figure 3.** Normalized LFN $WLS_{ID}f/I_D^2$, referred to $1$ Hz, vs. top gate voltage overdrive $V_{GEFF}$, for GFETs with a) $W/L=20$ μm/20 μm, b) $50$ μm/50 μm and c) $100$ μm/100 μm. Measured noise from all available samples: star markers, measured ln-mean noise and its ±2-sigma deviation: open circle markers, mean and ±2-sigma deviation model: lines. (mean data and model: black, +2-sigma deviation data and model: purple, -2-sigma deviation data and model: green)

$C=C_{top}+C_{back}$ is the sum of top and back interface capacitances, $\alpha_H$ is the unitless Hooge parameter which is also used as a LFN mean value model parameter[53] and $f$ is the frequency (see Section S.1 and Figure S1 in the Supporting Information for more details on the definition of different quantities.) To obtain the total LFN PSDs, integration of the local PSDs derived above with an integral variable change from length $x$ to chemical potential $V_c$ should take place.[53-55] The SG-SL GFETs which are examined in this study are long-channel devices and thus the long channel LFN mean value model proposed in Reference [53] is taken into account and not the more complicated one with the updates regarding Velocity Saturation effect on LFN[54] which will make the variance derivation too complicated. LFN mean value model is validated with experimental data, averaged in a bandwidth $1$-$30$ Hz, in **Figure 3** for all the GFETs under test. $WLS_{ID}f/I_D^2$ ln-mean data with black circular markers and LFN mean value model[53] with solid black lines are shown for the $W/L=20$ μm/20 μm GFET in Figure 3a, for the $W/L=50$ μm/50 μm GFET in Figure 3b and for the $W/L=100$ μm/100 μm GFET in Figure 3c vs. $V_{GEFF}$ with very consistent results for all the devices and for all regions of operation. The logarithmic (ln)-mean values are used for better accuracy due to log-normal distribution of LFN. The LFN data for the total of the measured samples are also shown with smaller red markers. $\Delta\mu$ model is significant near the CNP and from there the $\alpha_H$ parameter can be extracted, $\Delta N$ LFN is responsible for the M-shape of $WLS_{ID}f/I_D^2$ and thus, $N_T$ parameter can be extracted from fitting the $\Delta N$ model of LFN generation. Finally, the contribution from $\Delta R$ can be identified at higher gate voltage values, where $S_{\Delta R}^2$ can be extracted.[55] The LFN mean value model parameters are presented in Table 1. $N_T$ and $\alpha_H$ increase with the area of the GFETs, with the greatest increment observed in the largest devices ($W/L=100$ μm/100 μm), where they double with respect to the medium sized devices ($W/L=50$ μm/50 μm). Oppositely, $S_{\Delta R}^2$ reduces with the GFET channel area. The reason for this deviation of LFN mean value parameters is not well understood but GFET technologies are not as mature as CMOS for example, and this might cause such deviations even for parameters of the same technology. The ensuing LFN statistical analysis will confirm that $\Delta N$ and $\Delta\mu$ models define LFN variance similarly as its mean value.

*LFN Variance model:* The mechanisms that generate LFN are also accountable for its deviation.[39, 40] Variations in the number of active traps ($N_{tr}$) can definitely induce variations in the $\Delta N$ contribution to LFN. Similarly, variations in the $\alpha_H$ parameter can produce variability in the effect of $\Delta\mu$ on LFN. In this section, an analytical LFN variance model for each of the two LFN generation mechanisms will be derived, following a procedure similar to the LFN mean value model derivation.[53] First, LFN variance will be calculated locally in the channel and since the local noise sources are considered uncorrelated,[33, 53] integration from Source to Drain can provide the total LFN variance by adding all the local contributions. Estimating variance locally in the channel ensures that each $I_D$ deviation caused by a fluctuation (such as a specific trap) will contribute independently.[37, 39, 40] Oppositely, under uniform channel conditions the variance would be constant at every point of the channel and thus, it could be taken out of the integral. Fundamental statistics theory gives:

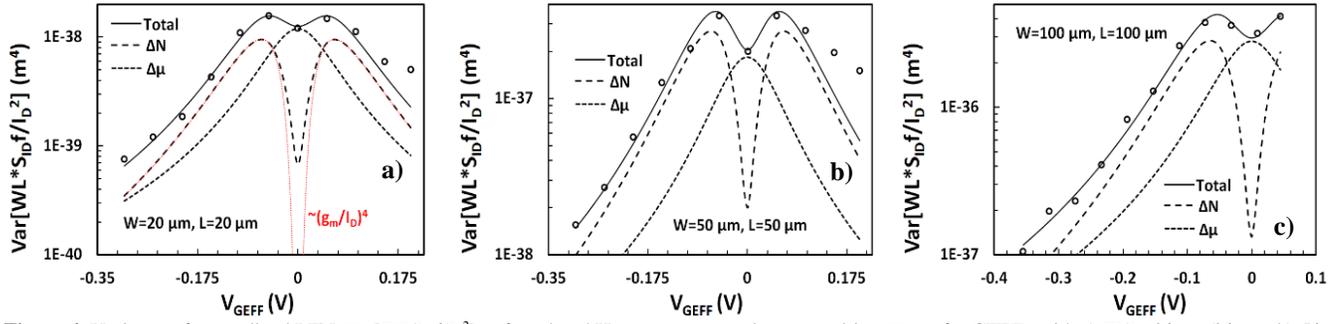

**Figure 4.** Variance of normalized LFN $Var[WLS_{ID}f/I_D^2]$, referred to *1* Hz, vs. top gate voltage overdrive $V_{GEFF}$, for GFETs with a) $W/L=20$ μm/20 μm, b) *50* μm/*50* μm and c) *100* μm/*100* μm. Markers: measured data, solid lines: total model, dashed lines: individual contributions ($\Delta N$, $\Delta \mu$). Simplified LFN variance model $\sim(g_m/I_D)^4$ which considers a homogeneous channel is shown with red dashed lines for $W/L=20$ μm/20 μm GFET (a) for comparison reasons.

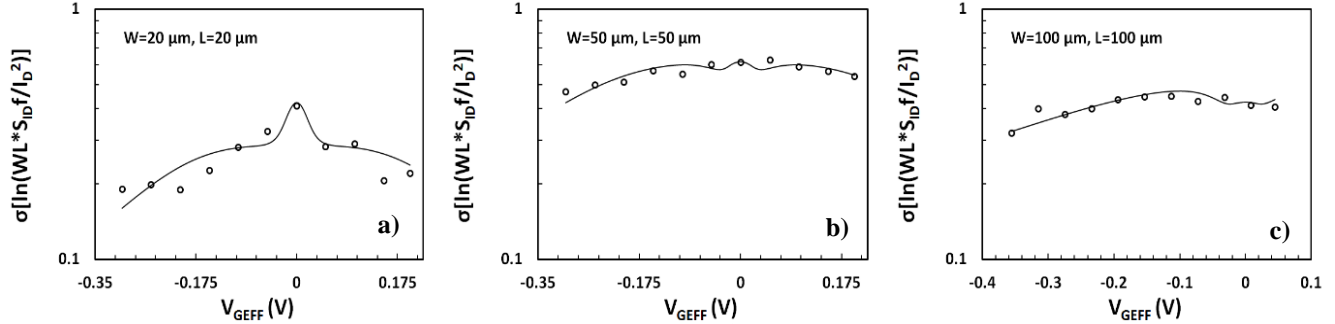

**Figure 5.** Standard deviation of natural logarithm of normalized LFN $\sigma[ln(WLS_{ID}f/I_D^2)]$, referred to *1* Hz, vs. top gate voltage overdrive $V_{GEFF}$, for GFETs with a) $W/L=20$ μm/20 μm, b) *50* μm/*50* μm and c) *100* μm/*100* μm. Markers: measured data, solid lines: total model.

$$Var\left[\frac{S_{I_D}}{I_D^2}\Bigg|_{\Delta N} WLf\right] = \frac{16kT\lambda N_{tcoeff}N_T e^4}{WLg_m C} \frac{\sqrt{k}}{8\left(C^2+ak\right)^5} \left[\begin{array}{l} -\text{sgn}\left(V_c\right)\dfrac{4C^4\sqrt{k}\left(C^2+ak\right)}{\left(C+kV_c^2\right)^3}-\text{sgn}\left(V_c\right)\dfrac{16C^3\sqrt{k}\left(C^2-2ak\right)\left(C^2+ak\right)}{C+\text{sgn}\left(V_c\right)kV_c}-\text{sgn}\left(V_c\right) \\[4mm] \dfrac{2a\sqrt{k}\left(C^2+ak\right)^2\left(a^3k-\text{sgn}\left(V_c\right)C^3V_c-3aC\left(C-\text{sgn}\left(V_c\right)kV_c\right)\right)}{\left(a+kV_c^2\right)^2}- \\[4mm] \dfrac{\sqrt{k}\left(C^2+ak\right)\left(\begin{array}{l}5C^3V_c+\text{sgn}\left(V_c\right)2aC^2\left(12C-\text{sgn}\left(V_c\right)17kV_c\right)\\+3aC^2k\left(-\text{sgn}\left(V_c\right)8C+3kV_c\right)\end{array}\right)}{a+kV_c^2} \\[4mm] +\dfrac{3C}{\sqrt{a}}\left(C^6-25\alpha C^4k+35a^2C^2k^2-3a^3k^3\right)\arctan\left(\sqrt{\dfrac{k}{a}}V_c\right)+\text{sgn}\left(V_c\right) \\[4mm] 24\sqrt{k}\left(C^6-5aC^4k+2a^2C^2k^2\right)\log\left(C+\text{sgn}\left(V_c\right)kV_c\right)- \\[4mm] \text{sgn}\left(V_c\right)12\sqrt{k}\left(C^6-5aC^4k+2a^2C^2k^2\right)\log\left(a+kV_c^2\right) \end{array}\right]_{V_{cs}}^{V_{cs}} \tag{9}$$

$$Var\left(f\left(y\right)\right)=\left[\frac{\partial f}{\partial y}\right]^2 Var\left(y\right) \tag{3}$$

where in this study $y=N_{tr}$, $\alpha_H$ for $\Delta N$, $\Delta \mu$ models respectively and $f(N_{tr}, \alpha_H)=\Lambda(x)(N_{tr}, \alpha_H)$.

*$\Delta N$ LFN Variance*: The total $WLS_{ID}f/I_D^2$ $\Delta N$ is calculated by integrating Equation (1) along the channel:[33, 53]

$$\frac{S_{I_D}}{I_D^2}\Bigg|_{\Delta N} WLf=\int_0^L \frac{1}{L^2}\left(\frac{e}{Q_{gr}}\frac{C_q}{C_q+C}\right)^2 \frac{N_{tr}}{W} dx \tag{4}$$

To calculate the variance of Equation (4):

$$Var\left[\frac{S_{I_D}}{I_D^2}\Bigg|_{\Delta N} WLf\right]=Var\left[\int_0^L \frac{1}{L^2}\left(\frac{e}{Q_{gr}}\frac{C_q}{C_q+C}\right)^2 \frac{N_{tr}}{W} dx\right] \tag{5}$$

It can be easily proved that since the local noise sources are uncorrelated and thus independent, the following Equation is valid:

$$Var\left[\int_0^L f\left(y\right)dx\right]=L\int_0^L Var\left[f\left(y\right)\right]dx \tag{6}$$

(See Section S.3 of the Supporting Information for the mathematical proof of Equation (6).) Because of Equation (3) and (6), variance of Equation (5) can enter the integral as:

$$Var\left[\frac{S_{I_D}}{I_D^2}\Bigg|_{\Delta N} WLf\right]=\int_0^L L\left(\frac{1}{WL^2}\left(\frac{e}{Q_{gr}}\frac{C_q}{C_q+C}\right)^2\right)^2 Var\left[N_{tr}\right]dx \tag{7}$$

where the quantity inside the integral corresponds to the local variance. In silicon-oxide devices number of traps are known to follow Poisson distribution[27, 35-38] which means that its variance equals to its mean value, $Var[N_{tr}]=N_{tr}=WLKT\lambda N_T$. It will be shown later that this is not the case in SG-SL GFETs of this work but it rather is $Var[N_{tr}]=N_{tcoeff}WLKT\lambda N_T$, where $N_{tcoeff}$ is used as a fitting LFN variance model parameter. According to the latter and if the integral variable is changed from $x$ to $V_c$:[53-58]

$$Var\left[\frac{S_{I_D}}{I_D^2}\Big|_{\Delta N} WLf\right] = \int_0^L \frac{kT\lambda N_{tcoeff} N_T e^4}{WL^2} \frac{C_q^4}{\left(C_q+C\right)^4 Q_{gr}^2} dx =$$

$$\int_{V_{cd}}^{V_{cs}} \frac{16kT\lambda N_{tcoeff} N_T e^4}{WL g_{vc} C} \frac{V_c^4}{\left(V_c^2+a/k\right)^3 \left(k|V_c|+C\right)^3} dV_c \tag{8}$$

and if Equation (8) is solved analytically Equation (9) above is derived. $g_{vc}$ is a normalized drain current term,[53-58] $\alpha = 2\rho_0 e$ is a residual charge related expression[55-58] and $k$ is a coefficient.[53-58] (See also Section S.1 of Supporting Information.) Equation (9) predicts an inversely proportional relation of the $WLS_{ID}f/I_D^2$ $\Delta N$ variance model to the area of the device ~$1/(WL)$.

*$\Delta\mu$ LFN Variance*: Following a procedure similar as in $\Delta N$ case, the total $WLS_{ID}f/I_D^2$ $\Delta\mu$ variance is calculated by integrating Equation (2) along the channel:[33, 53]

$$\frac{S_{I_D}}{I_D^2}\Big|_{\Delta\mu} WLf = \int_0^L \frac{\alpha_H e}{LQ_{gr}} dx \tag{10}$$

Variance of Equation (10) is calculated after taking into consideration Equation (3) and (6) as in $\Delta N$ case since local noise sources are uncorrelated:

$$Var\left[\frac{S_{I_D}}{I_D^2} WLf\right]_{\Delta\mu} = \int_0^L \frac{e^2}{LQ_{gr}^2} Var\left[\alpha_H\right] dx \tag{11}$$

No information is available on literature regarding $Var[\alpha_H]$ and thus, in order to achieve a similar scaling with $\Delta N$ $WLS_{ID}f/I_D$ LFN variance model ~$1/(WL)$, the following is assumed: $Var[\alpha_H]=\alpha_H(N_{\alpha H}WL)^{-1}$ where $N_{\alpha H}$ is a specific density used as a fitting LFN variance model parameter. According to the latter and if the integral variable is changed from $x$ to $V_c$:[53-58]

$$Var\left[\frac{S_{I_D}}{I_D^2}\Big|_{\Delta\mu} WLf\right] = \frac{4e^2\alpha_H}{WLN_{\alpha_H}g_{vc}Ck^2}\int_{V_{cd}}^{V_{cs}} \frac{k|V_c|+C}{V_c^2+a/k} dV_c \tag{12}$$

and if Equation (12) is solved analytically:

$$Var\left[\frac{S_{I_D}}{I_D^2}\Big|_{\Delta\mu} WLf\right] = \frac{4e^2\alpha_H}{WLN_{\alpha_H}g_{vc}Ck}\left[\frac{C\arctan\left(\sqrt{\frac{k}{a}}V_c\right)}{\sqrt{ak}}+0.5\mathrm{sgn}(V_c)\left(a+kV_c^2\right)\right]_{V_{cd}}^{V_{cs}} \tag{13}$$

*Total LFN Variance*: The total $WLS_{ID}f/I_D^2$ variance can be calculated as:

$$Var\left[\frac{S_{I_D}}{I_D^2} WLf\right] = Var\left[\frac{S_{I_D}}{I_D^2}\Big|_{\Delta N} WLf\right] + Var\left[\frac{S_{I_D}}{I_D^2}\Big|_{\Delta\mu} WLf\right] \tag{14}$$

under the approximation that $\Delta N$ and $\Delta\mu$ models are uncorrelated. Even though this is not completely accurate, there are distinguished regions where each of these effects is dominant ($\Delta\mu$ at the CNP and $\Delta N$ at the peaks of M-shape dependence) and thus the aforementioned independency can be assumed without significant error. Equation (9), Equation (13) and Equation (14) formulate the new compact statistical LFN model.

As shown before, $WLS_{ID}f/I_D^2$ follows a log-normal distribution as it can be observed by the CDFs illustrated in Figure 2b and in Figure S2a, S2b of the Section S.2 of the Supporting Information. The fundamental Equation for this distribution is:[37, 39]

$$\sigma\left(\ln\left[\frac{S_{I_D}}{I_D^2} WLf\right]\right) = \sqrt{\ln\left(1+\frac{Var\left[\frac{S_{I_D}}{I_D^2} WLf\right]}{E^2\left[\frac{S_{I_D}}{I_D^2} WLf\right]}\right)} \tag{15}$$

Where $\sigma(ln[WLS_{ID}f/I_D^2])$ is the standard deviation of the natural logarithm of $WLS_{ID}f/I_D^2$ widely used in bibliography,[25, 27, 35-37, 39-40] $E$ denotes the LFN mean value model and the ratio of variance with squared mean $WLS_{ID}f/I_D^2$ is defined as normalized variance.[25, 27, 35-37, 39-40] The model in Equation (15) follows the scaling dependence of Reference [37], which is ~$\sqrt{ln[1+K/(WL)]}$ and for larger devices turns to ~$\sqrt{1/(WL)}$ where $K/(WL)$ is the normalized variance established before.[37]

*Experimental Validation of the LFN Variance Model*: The qualitative performance of the new derived LFN variance model is verified with the data from SG-SL GFETs under test, as it will be illustrated in the rest of this section. Initially, as it was mentioned before, ln-mean of $WLS_{ID}f/I_D^2$ data is calculated for every device under test from all the available samples and used in the verification of the LFN mean value model as in Figure 3. Afterwards, standard deviation $\sigma(ln[WLS_{ID}f/I_D^2])$ data can be easily extracted from the natural logarithms of all the $WLS_{ID}f/I_D^2$ samples again for each available GFET. This process is followed for the derivation of $\sigma(ln[WLS_{ID}f/I_D^2])$ instead of Equation (15) since noise measurements are very sensitive and thus the calculation of normalized variance ratio $Var[WLS_{ID}f/I_D^2]/E^2[WLS_{ID}f/I_D^2]$ contained in Equation (15) is not very consistent because of the small numbers both in numerator and denominator. Then, $WLS_{ID}f/I_D^2$ variance can be estimated through Equation (15) since $\sigma(ln[WLS_{ID}f/I_D^2])$ is already known, by using the ln-mean in the denominator of normalized variance. In Figure 3, $\pm2\sigma$ standard deviation of $WLS_{ID}f/I_D^2$ ($\sigma=\sqrt{Var}$) is also shown both for the model and experimental data with purple (+2σ) and green (-2σ) solid lines and markers respectively. The model captures accurately the dispersion of the data and its bias dependence, confirming the consistency between LFN mean value and variance models for all the three GFETs examined.

$WLS_{ID}f/I_D^2$ variance and $\sigma(ln[WLS_{ID}f/I_D^2])$ are depicted in **Figure 4** and **Figure 5** respectively for the $W/L=20$ μm/20 μm GFET in Figure 4a and 5a, for the $W/L=50$ μm/50 μm GFET in Figure 4b and 5b and for the $W/L=100$ μm/100 μm GFET in Figure 4c and 5c vs. $V_{GEFF}$. Experimental data are represented with markers while the total model with solid lines. Dashed and dotted lines in Figure 4 stand for the $\Delta N$ and $\Delta\mu$ variance contributions, respectively. This representation confirms that the model precisely captures the experimental data for both $WLS_{ID}f/I_D^2$ variance and $\sigma(ln[WLS_{ID}f/I_D^2])$ in the whole range of operation for every GFET under investigation.

The $\Delta N$ and $\Delta\mu$ models shown in Figure 4 prove that these effects act similarly as in the LFN mean value model, $\Delta N$ effect is responsible for the M-shape of $WLS_{IDf}/I_D^2$ variance as it was for the $WLS_{IDf}/I_D^2$ mean value while $\Delta\mu$ contributes near the CNP as it did for the LFN mean value model while it retains its $\Lambda$ shape trend. The new LFN variance model shows a deviation from the experimental data at high n-type conduction regime probably due to $\Delta R$ contribution which is not included in this study. For comparison reasons, a LFN variance model based on $\sim(g_m/I_D)^2$ approximation which considers a uniform channel,[26, 29, 31] is shown in Figure 4a with red dashed line. This approach estimates an $\sim(g_m/I_D)^4$ dependence of LFN variance. (see Section S.4 of the Supporting Information for the analysis of the derivation of this expression.) As it was expected, it gives acceptable results away from the CNP but it is incapable of capturing LFN variance near the CNP where the non-homogeneity of the device is more intense even for small $V_{DS}$ values.[53] These non-homogeneities can be easily detected from the illustration of the $\Delta N$ LFN contribution to $WLS_{IDf}/I_D^2$ variance throughout the channel where, for $V_{GEFF}$ values away from the CNP, local variance is constant along the channel. Oppositely, near the CNP a steep dip is noticed in the middle of the channel in which the exact CNP is located under low $V_{DS}$. (For the above observations see Figure S3 in Section S.5 of the Supporting Information.) The experimental data were measured at $V_{DS}=50$ mV, but it is quite certain that the inconsistency of the $\sim(g_m/I_D)^4$ term would be more significant for higher $V_{DS}$ values.

$N_{tcoeff}$, $N_{aH}$ parameters of $\Delta N$ and $\Delta\mu$ effects respectively, are extracted and shown in Table 1. $N_{aH}$ is calculated from the CNP while $N_{tcoeff}$ is then adjusted to fit the M-shape. It is clear that $N_{tcoeff}$ values for all GFETs are far from unity which means that $N_{tr}$ does not follow a Poisson distribution as in silicon-oxide devices. As stated before, this might be due to the nature of the traps in graphene-electrolyte interface[43, 60-63] of the specific SG-SL GFETs. In order to prove the validity of the obtained $N_{tcoeff}$, $N_{aH}$ values, a thorough analysis was conducted where the LFN mean value model parameters ($N_T$-$N_{tr}$, $\alpha_H$) were extracted for each of the measured samples for all available GFET areas. The variance of these parameters as well as their ln-mean value were then estimated, allowing to derive the LFN variance model parameters $N_{tcoeff}$, $N\alpha_H$, given that $Var[N_{tr}]=N_{tcoeff}WLKT\lambda N_T$ and $Var[\alpha_H]=\alpha_H(N\alpha_H WL)^{-1}$. These values are proven to be identical with the ones extracted from Figure 4 and shown in Table 1, which provide the best possible fitting of the model. This result confirms that charge traps do not present a Poisson distribution in the GFETs under study. (For the complete analysis see Figure S4 and Table S1 in the Section S.6 of the Supporting Information.) As it was mentioned before, $WLS_{IDf}/I_D^2$ mean value increases as the area gets larger and as a result $N_T$, $\alpha_H$ get also higher as shown in Table 1. Table S1 clearly demonstrates that variances of $N_{tr}$, $\alpha_H$ respectively, increase more strongly than their mean values as the devices' area increases. The latter can justify the boost of $N_{tcoeff}$ parameter and the reduction of $N_{aH}$ as the dimensions get larger. The reason for these variations of the statistical LFN model parameters which occur as a physical consequence of the variations of the LFN mean value model ones, could be the critical inhomogeneities of GFET technologies, as it has already been stated before.

Standard deviation of natural logarithm $\sigma(ln[WLS_{IDf}/I_D^2])$ of normalized LFN is shown in Figure 5 and some remarkable conclusions can be extracted. The above quantity can be used as a figure of merit for LFN variability comparisons between the GFET technology in this work and CMOS ones[35-37, 40] It can be concluded that the range of values between $0.1$-$1$ for all the transistors under test as depicted in Figure 5 are similar with the results obtained from CMOS devices with similar dimensions, indicating a decent performance for the GFETs under study.[35-37, 40] Another crucial observation is the weak bias dependence of $\sigma(ln[WLS_{IDf}/I_D^2])$ data with a rather smooth fluctuation near the CNP and a slight decrease away from the CNP, which is remarkably captured by the proposed model.

Despite the fact that the proposed $WLS_{IDf}/I_D^2$ variance model focuses on the bias dependence, its scaling with the area is also crucial. $\Delta N$ and $\Delta\mu$ $WLS_{IDf}/I_D^2$ variance models follow a $\sim 1/(WL)$ trend as it is clear from Equation 9 and 13 and as a result total LFN variance expression in Equation (14) behaves similarly. $S_{IDf}/I_D^2$ mean value model presents the same dependency.[53-55] In addition, $\sigma(ln[WLS_{IDf}/I_D^2])$ also follows an $\sim\sqrt{1/(WL)}$ trend due to the large device dimensions.[25, 27, 35-37, 40] However, due to inhomogeneities of the GFET technologies in general, extracted parameters both of the LFN mean value ($N_T$, $\alpha_H$) and variance ($N_{tcoeff}$, $N_{aH}$) models, differ from device to device.

## III. CONCLUSIONS

This work investigates thoroughly the bias dependence of LFN variability in large area SG-SL GFETs. This the first time that such an analysis of statistical LFN data for graphene devices is presented. An analytical compact model based on carrier number $\Delta N$ and mobility fluctuation $\Delta\mu$ effects has been proposed and implemented for circuit simulators. The development of such a model is critical for the boost of graphene circuit design, where LFN variability should be accurately predicted to prevent performance deterioration in certain applications. In this work, it is experimentally proven that LFN variability does not count on the variability of IV quantities such as $V_{CNP}$, but it is directly linked to the number of traps $N_{tr}$ and Hooge parameter $\alpha_H$ variations regarding $\Delta N$ and $\Delta\mu$ mechanisms respectively. The derived compact model precisely covers the measured LFN variance over the whole range of operation, from strong conduction to the CNP at both p- and n-type regimes. $\Delta N$ and $\Delta\mu$ models determine LFN variance in the same way as its mean value. Thus, $\Delta N$ effect accounts for the M-shape of LFN variance similarly as it is known to cause an M-shape for its mean value while $\Delta\mu$ provides a $\Lambda$-shape in the bias dependence of LFN variance, which contributes significantly at the CNP, analogously to the $\Delta\mu$ contribution to LFN mean value. A simpler variance model with a ~$(g_m/I_D)^4$ shape is also extracted based on a uniform channel approximation and shown for comparison reasons. This approach based on the well known ~$(g_m/I_D)^2$ model of the $\Delta N$ contribution to LFN mean value fails to accurately predict the LFN variance near the CNP.

$N_T$, (or $N_{tr}$ consequently) and $\alpha_H$ parameters of the LFN model mean value are also used in LFN variance model together with the newly defined $N_{tcoeff}$ and $N_{\alpha H}$ parameters which are extracted from statistical LFN data. In silicon-oxide transistors, $N_{tr}$ follows a Poisson distribution and thus, $N_{tcoeff}$ is close to unity but this is not the case in the devices' under test. In these SG-SL GFETs, apart from the traps in the polyimide substrate beneath the graphene, an additional different type of traps is present in the graphene-electrolyte interface which might be responsible for the non-Poisson distribution of $N_{tr}$. The latter is experimentally shown by extracting LFN mean value model parameters for every measured sample and then by calculating their variance and mean value.

Regarding geometrical scaling, both $S_{ID}f/I_D^2$ mean value and $WLS_{ID}f/I_D^2$ variance models follow a ~$1/(WL)$ behavior while standard deviation of natural logarithm of $WLS_{ID}f/I_D^2$ $\sigma(ln[WLS_{ID}f/I_D^2])$ a ~$\sqrt{1/(WL)}$ shape for larger devices which is the case in the present study. The latter is taken into consideration and examined due to log-normal distribution of LFN data and it might be a reliable tool in order to compare LFN variance for different types of transistors. In order the above area dependencies to be valid, identical LFN parameters are needed for devices of the same technology. This is not the case for the GFETs of this work since the LFN mean value parameters increase with the device area causing also a large deviation in LFN variance parameters. This device to device parameters' deviation could be explained by increased inhomogeneities observed on recently developed GFET technologies. This study represents the first reported efforts to understand LFN variability in GFETs. The derived results contribute to the thorough understanding of the nature of charge traps statistics in solution-gated devices and they also provide the tools to quantify and predict LFN variability in SL GFETs. This framework is considered critical for upscaling the production of graphene electronics from research labs into larger-scale dedicated fabrication facilities.

## IV. EXPERIMENTAL SECTION

*Electrical characterization of LFN*: To measure the DC transfer curves and the LFN spectra accurately, the drain to source current was pre-amplified in a first amplification stage with a gain of $10^4$. The signal was then high-pass filtered, thus canceling its low frequency (i.e. DC level) components. The resulting signal was further amplified and low-pass (anti-aliasing) filtered in a second stage with a $10^2$ gain. The signals were digitalized using a NI DAQ Card in all characterization procedures. To extract the power spectral density, the drain to source current was measured under different gate bias conditions for $10$ s at each point.

*Fabrication of SG GFETs*: Arrays of SG-SL GFETs were fabricated on a $10$ μm thick polyimide (PI-2611, HD MicroSystems) film spin coated on a Si/SiO2 $4$" wafer and baked at $350$°C. A first metal layer ($10$ nm Ti/$100$ nm Au) was deposited by electron-beam vapour and then structured by a lift-off process. Afterwards, the graphene grown by chemical vapour deposition on Cu was transferred (process done by Graphenea s.a.). Graphene was then patterned by oxygen plasma ($50$ sccm, $300$ W for $1$ min) in a reactive ion etching (RIE) after protecting the graphene in the channel region with HIPR 6512 (FujiFilm) positive photoresist. After the graphene etching, a second metal layer was patterned on the contacts following the same procedure as for the first layer. The lift-off was followed by an annealing in ultra-high vacuum consisting on a temperature ramp from room temperature to $300$ °C. Subsequently, the transistors were insulated with a $3$-μm-thick photodefinable SU-8 epoxy photoresist (SU-8 2005 Microchem), keeping uncovered the active area of the transistors channel and the contacting pads. The polyimide substrate was structured in a reactive ion etching process using a thick AZ9260 positive photoresist (Clariant) layer as an etching mask. The neural probes were then peeled off from the wafer and placed in a zero-insertion force connector to be interfaced with our custom electronic instrumentation. Finally, the devices were rinsed for $2$ minutes in ethanol to eliminate remaining resist residues on the graphene channel.

*Raman characterization of graphene after transfer*: A SLG sample was transfer onto a SiO2 wafer following the same process as detailed for the fabrication of GFETs. The Raman spectra at $400$ equally spaced points were acquired on the graphene sample, within an area of $20$ μm x $20$ μm. A Witec spectrometer in backscattering configuration, using a $600$ gr/nm grating was used. A $488$ nm wavelength laser ($2.5$ mW power) was focused on the sample with a 50x objective. The peak intensity for the D and G bands was measured after background subtraction.

TABLE I
IV-LFN (MEAN-VARIANCE) MODEL EXTRACTED PARAMETERS

| Parameter | Units | 20/20 | 50/50 | 100/100 |
|---|---|---|---|---|
| $\mu$ | cm$^2$(Vs)$^{-1}$ | 3500 | 3800 | 7000 |
| $C_{ox}$ | μFcm$^2$ | 2 | 2 | 2 |
| $V_{GSO}$ | V | 0,278 | 0,277 | 0,33 |
| $\rho_0$ | cm$^2$ | 3.29.10$^{11}$ | 3.29.10$^{11}$ | 3.43.10$^{11}$ |
| $R_c/2=R_{S,D}$ | Ω | 400 | 320 | 380 |
| $N_T$ | eV$^{-1}$cm$^{-3}$ | 2·2.10$^{19}$ | 4.93·10$^{19}$ | 1.74·10$^{20}$ |
| $\alpha_H$ | - | 5.6·10$^{-4}$ | 1.5·10$^{-3}$ | 1.12.10$^{-2}$ |
| $S_{\alpha R}^2$ | Ω$^2$/Hz | 3·10$^{-4}$ | 4·10$^{-5}$ | 1·8.10$^{-5}$ |
| $N_{tcoeff}$ | - | 2.7.10$^5$ | 2.5.10$^5$ | 3.4.10$^6$ |
| $N_{\alpha H}$ | - | 1.9.10$^{13}$ | 3.5.10$^{11}$ | 3.6.10$^{10}$ |

## ACKNOWLEDGEMENTS

This work was funded by the European Union's Horizon 2020 research and innovation program under Grant Agreement No. GrapheneCore2 785219 and No. GrapheneCore3 881603, Marie Skłodowska-Curie Grant Agreement No 665919 and Grant Agreement No. 732032 (BrainCom). We also acknowledge financial support by Spanish government under the projects TEC2015-67462-C2-1-R, RTI2018-097876-B-C21 (MCIU/AEI/FEDER, UE) and project 001-P-0011702-GraphCat: Communitat Emergent de grafè a Catalunya, co-funded by FEDER within the framework of Programa Operatiu FEDER de Catalunya 2014-2020. The ICN2 is also supported by the Severo Ochoa Centres of Excellence programme, funded by the Spanish Research Agency (AEI, grant no. SEV-2017-0706).